# Early stage of CVD graphene synthesis on Ge(001) substrate

L. Di Gaspare*[1], A. M. Scaparro[1], M. Fanfoni[2], L. Fazi[2], A. Sgarlata[2], A. Notargiacomo[3], V. Miseikis[4], C. Coletti[4], and M. De Seta[1]

[1]*Dipartimento di Scienze, Università Roma Tre, Viale G. Marconi, 446- 00146 Rome, Italy*
[2]*Dipartimento di Fisica, Università di Roma "Tor Vergata," Via della Ricerca Scientifica, 1-00133 Roma, Italy*
[3]*Institute for Photonics and Nanotechnology, CNR, Via Cineto Romano 42, 00156 Rome, Italy*
[4]*Center for Nanotechnology Innovation @NEST, Italian Institute of Technology, Piazza San Silvestro 12, 56127 Pisa, Italy*

**Abstract**

In this work we shed light on the early stage of the chemical vapor deposition of graphene on Ge(001) surfaces. By a combined use of $\mu$-Raman and x-ray photoelectron spectroscopies, and scanning tunneling microscopy and spectroscopy, we were able to individuate a carbon precursor phase to graphene nucleation which coexists with small graphene domains. This precursor phase is made of C aggregates with different size, shape and local ordering which are not fully $sp^2$ hybridized. In some atomic size regions these aggregates show a linear arrangement of atoms as well as the first signature of the hexagonal structure of graphene. The carbon precursor phase evolves in graphene domains through an ordering process, associated to a re-arrangement of the Ge surface morphology. This surface structuring represents the embryo stage of the hills-and-valleys faceting featured by the Ge(001) surface for longer deposition times, when the graphene domains coalesce to form a single layer graphene film.

* Corresponding author. Tel : +39 06 5733 3315 E- mail : : <u>luciana.digaspare@uniroma3.it</u> (Luciana Di Gaspare)

# 1. Introduction

Catalyzed chemical vapor deposition (CVD) on metallic substrates has been largely predicted as one of the most promising techniques for the scalable synthesis of highly crystalline graphene, which is necessary for the development of graphene based electronics [1]. However, the graphene integration in standard complementary metal oxide semiconductor (CMOS) technology is hindered by metallic impurities and defects which are introduced by the growth process itself [2] or successively in the transfer process on Si wafers [3-4]. A significant improvement toward the compatibility of CVD graphene with CMOS-technology is represented by the recent achievement of metal contamination-free graphene grown directly on Ge or Ge/Si substrates [5-6], in particular on the technology relevant (001) surface orientation [7-13]. Despite this remarkable breakthrough, the quality of graphene deposited on Ge(001) should still be improved for "real-world" technological applications. To this end, it is necessary to investigate, understand, and acquire a full control over the adsorption and nucleation mechanisms of the carbon species on the substrate at the early stage of graphene growth, which deeply influence the quality of the graphene at all the subsequent stages of the deposition.

Theoretical studies of hydrocarbon decomposition mechanisms and nucleation on metal surfaces (such as Ir, Cu, Ni, Ru), have focused on competing roles of the C-C and carbon-metal interaction at the graphene-substrate interface on graphene nucleation [14-18]. In different substrates, this competition leads to different structures acting as stable precursors [14, 16, 19]. Early stage of graphene synthesis on metal substrates (e.g. Cu, Ir) was experimentally investigated in Refs. [20-27]. A binding between the graphene precursor phase and the substrates has been evidenced by x-ray photoelectron spectroscopy (XPS) [21,27]. Gao et al. reported that precursors of the graphene growth on Ru were made of chains of C dimers [20], while in Ref. [24] was reported that graphene nuclei expand their lateral sizes by the addition of clusters comprising ~5 C atoms. Concerning the graphene-Cu system, whose similarity with the graphene-Ge one has been recently evidenced in [10], different carbon clusters were identified by scanning tunneling microscopy (STM) as "growth intermediates" prior to the graphene formation and their evolution after the saturation of the surface in defected graphene was observed [22].

As for the CVD growth on Ge(001) substrate, the first stage of graphene nucleation has not yet been clarified yet. The quality and characteristics of the deposited graphene depend strongly on the Ge surface orientation [5,6,8,10,11], indicating that the C atom interaction with the Ge surface is of paramount importance in determining the graphene quality. Although Ge forms no stable carbide phase (GeC), theoretical calculations [28] suggest that, in graphene growth from solid C sources, the interaction between the Ge(001) surface and C atoms results in the immobilization of a C atom by substitution of a Ge atom in a surface dimer, so that either carbon dimers trapped by Ge dimer vacancies or longer C chains trapped between dimer rows act as graphene seed. As for CVD graphene, the same study [28] predicts that $CH_x$ diffusing species can react with one another leading to the formation of polymeric carbon rings stabilized on the Ge(001) by GeC bonds.

In this work we experimentally investigate the early stage of CVD graphene synthesis on Ge(001) substrate using methane as precursor gas. To this end, we combine spectroscopic measurements (Raman spectroscopy and x-ray photoelectron spectroscopy) and an atomic scale characterization by means of scanning tunneling microscopy. Our analyses reveal that at the early stage of graphene growth a carbon precursor phase still not organized in the hexagonal atomic structure of graphene is present and covers a large part of the Ge surface. This C precursor phase evolves in graphene nuclei that in turn coalesce to form a single layer graphene (SLG).

## 2. Materials and methods

Graphene was deposited on Ge(001) substrates in a commercial 4-inches CVD system (BM, Aixtron) using $CH_4$ and $H_2$ as precursor gases and Ar as a carrier gas. $CH_4$, $H_2$, and Ar fluxes were set at 2, 200, and 800 sccm, respectively. The growth pressure was 100 mbar and the substrate temperature was fixed at 930 °C. In these conditions, the graphene grows following a layer by layer regime in which the second graphene layer starts to develop only after the completion of the first one [10]. The substrate heating was carried out by a multi-step temperature ramp that ensures good reproducibility and a homogeneous temperature on the whole Ge surface maintaining a proper surface morphology up to 930 °C, i.e. few degrees below the Ge surface melting.

The sample structure and morphology were characterized by Raman spectroscopy, XPS, STM and scanning tunneling spectroscopy (STS). Raman spectroscopy (Renishaw inVia confocal Raman microscope) was performed using an excitation wavelength of 532 nm, a 100× objective, resulting in a laser spot size of ~1 $\mu$m. In order to compare the emission intensities of 2D, G, and D bands between different samples the Raman data were acquired in the same experimental conditions and we also checked that the emission from $N_2$ at 2331 $cm^{-1}$ coming from aerial contamination had the same intensity in all the spectra. The intensity of the 2D, G and D bands were evaluated by using the integrated peak area. The XPS measurements were carried out using a monochromatic Al $K_\alpha$ source (hv=1486.6 eV) and a concentric hemispherical analyzer operating in retarding mode (Physical Electronics Instruments PHI), with overall resolution of 0.35 eV.

The carbon surface coverage ρ was evaluated by using the C1s core level area intensity of each sample normalized to that acquired in the same experimental condition on a commercial graphene monolayer (CGM) (i.e. graphene grown via CVD on copper foil and transferred on $SiO_2$ [10]) mounted on the sample holder beside the analyzing sample, i.e. ρ=$I_{C1s}$(sample)/$I_{C1s}$(CGM)=$N_c$(sample)/$N_c$(CGM), where $N_c$ is the number of carbon atoms for unit area. In particular, the sample grown for 60 min has a surface coverage ρ=0.97±0.05 corresponding, within the experimental error, to a single layer graphene.

The STM/STS measurements were carried out under ambient conditions at room temperature employing a WA technology TOPS System. Electrochemically etched W tips were used for STM/STS measurements. Topographic images were acquired in constant current mode.

## 3. Results

*3.1 Raman and x-ray photoelectron spectroscopy measurements*

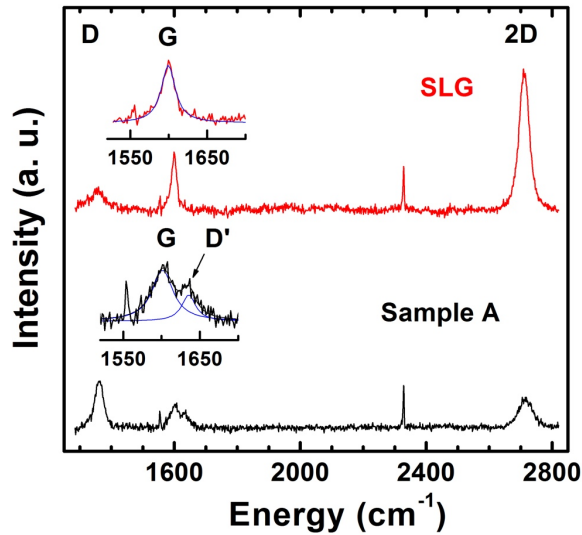

*Figure 1.* *Raman spectra of sample A (black line, deposition time 30 min) and SLG (red line, deposition time 60 min). The insets show in more detail the G bands at 1600 $cm^{-1}$ evidencing the presence of the D' peak in the spectrum of sample A. Blue lines are the Lorentzian fit of the G and D' peaks. The peaks at ~1554 and 2331 $cm^{-1}$ are due to $O_2$ and $N_2$ respectively.*

|  | $E_{2D}(cm^{-1})$ | $E_G(cm^{-1})$ | $\Gamma_{2D}(cm^{-1})$ | $I_D/I_{D'}$ |
|---|---|---|---|---|
| Sample A | 2712.7±1.7 | 1601.3±1.3 | 50.3 | 4.7±0.8 |
| SLG | 2711.3±3.5 | 1601.5±3 | 36 | - |

*Table 1. 2D and G peak energies, 2D FWHM ($\Gamma_{2D}$) and $I_D/I_{D'}$ intensity ratio extracted from the Raman spectra of sample A and SLG. The values are averaged on several spectra acquired on different surface regions. The reported errors represent half of the maximum dispersion of the experimental values.*

To gain insight into the early stage of CVD graphene synthesis, we deposited graphene films varying the deposition time $t_D$. In Figure 1 the Raman spectrum of a sample grown for $t_D$=30 min (sample A, black line) is compared to that of a single layer graphene covering uniformly the Ge surface and deposited using the same growth conditions for 60 min (SLG sample, red line). The main parameters obtained from the Raman analysis are reported in Table 1. Significant differences can be observed between the two spectra. The main graphene features, i.e. 2D and G peaks, are present in both the samples. The SLG has a narrow 2D Lorentzian peak with a FWHM $\Gamma_{2D}$ of 36 cm$^{-1}$ and an integrated $I_{2D}/I_G$ intensity ratio equal to 3.6. At about 1300 cm$^{-1}$ a residual D peak (integrated $I_D/I_G$ intensity ratio ~0.3) related to intervalley resonant scattering induced by defects is also visible. The Raman spectrum of sample A presents the typical signature of disorder [29,30] and therefore may be characteristic of graphene films at the very early stage of the growth. The spectrum is indeed characterized by a larger D peak and less intense and wider 2D peak. Since the graphene samples are grown in a layer by layer regime, no contribution to the 2D width from multilayer graphene domains are expected. The increase of the $\Gamma_{2D}$ could be due to strain variation on length scale below the laser spot size as suggested in [31]. Despite the significantly different $\Gamma_{2D}$ values, the averaged $E_{2D}$ and $E_G$ energy positions of sample A and SLG are very similar as shown in Table 1. This suggests that the average strain and doping level of sample A are similar to those of SLG in which a charge density of the order of $1 \times 10^{13}$ cm$^{-2}$ and a compressive biaxial strain of the order of $\varepsilon \approx -0.3\%$ [10] were found following the method reported in Ref. [32]. As shown by the inset, in sample A at 1635 cm$^{-1}$ also the D' peak, related to intravalley resonant scattering process induced by defects, is visible. Thus the G peak region was fitted with two Lorentzian curves corresponding to G and D' bands while a single Lorentzian component was used for the SLG. The value of the $I_D/I_{D'}$ intensity ratio gives information on the nature of defects present in graphene. Experimental $I_D/I_{D'}$ values measured on intentionally defected graphene are ~13 for on-site defects (sp$^3$ defective C atom clusters of 20-30 nm in size), ~7 for hopping defects (which represent vacancies and defects that produce deformation of C-C bonds in graphene) and ~3.5 for grain boundary defects measured in graphite [33]. In our sample A the $I_D/I_{D'}$ ratio is equal to 4.7 suggesting that vacancy and grain boundary are the dominant defects. Notice that ab initio calculations performed assuming isolated defects predict a $I_D/I_{D'}$ ratio about 10 for hopping defects and of 1.3 for on site defects [34]. Therefore the presence of isolated sp$^3$ defects cannot be excluded.

In Figure 2 the high resolution C1s and Ge3d XPS spectra of the two samples are reported. As displayed in panel (a), the SLG shows a C1s spectrum with the typical graphene asymmetric lineshape peaked at 284.4 eV well fitted using the Doniach-Sunjic profile (asymmetry parameter of 0.12). The absence of a C1s component at lower energy indicates that Ge–C bonds are negligible. The C1s spectrum of sample A has a lower intensity corresponding to a surface coverage ρ=0.75. Furthermore, the C1s peak is shifted at a lower binding energy and exhibits a different lineshape well fitted by two components centered at 284.4 eV and 284.1 eV respectively (lower curves of Figure 2 (a)).

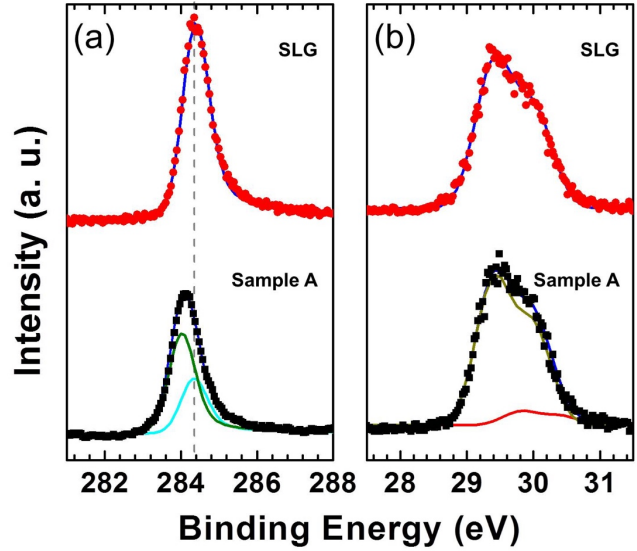

***Figure 2.*** *High resolution C1s and Ge3d XPS spectra of sample A (black symbols) and SLG (red symbols). (a): C1s spectra. Upper curve: the SLG spectrum was fitted with Doniach-Sunjic profile (blue line). Lower curve: in the sample A the fitting curve (blue line) is the sum of a Doniach-Sunjic component centered at 284.4 eV for the $sp^2$ GD phase (cyan line) and a gaussian component at 284.1 eV per the CP phase (green line). (b): Ge3d spectra. Upper curve: the SLG spectrum was fitted with a $3d_{5/2}$ and $3d_{3/2}$ spin-orbit splitting doublet (blue line) corresponding to the Ge-Ge bond. Lower curves: the 3d spectrum of the sample A was fitted with the sum of two spin orbit splitting doublets corresponding to the Ge-Ge bond (green line) and Ge- C bond (red line). The blue line is the obtained fit.*

In sample A, the C1s components at 284.4 eV has the same shape of the SLG C1s spectrum suggesting the presence of graphene domains (GD phase). Its intensity is the 35% of the whole spectrum, corresponding to a coverage of 0.25. The additional C1s component at 284.1 eV has a Gaussian lineshape and accounts for the 65% of the intensity of the whole spectrum corresponding to a coverage equal to 0.5. This component is not related to contamination or adventitious carbon: as a matter of fact in literature it is reported that C-H bonds were located at 284.5 eV, alcohol C-OH (or C-O-C) bonds at 286 eV and carbonyl (C=O) bonds at 288.4 eV [35]. Also in Ge substrate submitted to the growth procedure reported in this paper but without methane in the gas mixture (where C can only be present due to contamination), and in uncleaned pristine Ge substrate the C1s emission is located at higher binding energy [10,36]. The C1s component at 284.1 eV is therefore due to an extra C phase developing on the Ge surface in the first stage of the CVD growth. In the following we will refer to this phase as carbon precursor phase (CP phase). The Ge3d core level spectra are reported in Figure 2(b). The SLG spectrum is well fitted with a single spin-orbit doublet with $3d_{5/2}$ binding energy equal to 29.4 eV, $3d_{5/2}$-$3d_{3/2}$ energy splitting of 0.6 eV and intensity ratio of 0.66, respectively. Since the Ge-C bond is negligible in this sample this spectrum is related to Ge atoms bonded to Ge. The best fit of the Ge3d spectrum of sample A suggests the presence of two spin- orbit doublet components with $3d_{5/2}$ peak centered at 29.4 and 29.8eV, respectively. The doublet at 29.4 eV has the same lineshape and energy position of the SLG one and corresponds to the signal coming from Ge atoms bonded only to Ge. As for the doublet at higher binding energy, it is likely due to bonds between Ge surface atoms and C atoms of the grown film. Its binding energy is compatible with Ge atoms with one or at most two Ge-C bonds [37] being the energy chemical shift lower than the value reported for GeC bond in carbide form [38]. Since the electron escape depth of Ge3d photoelectrons in Ge is $\lambda_{Ge}\sim 2$ nm [39], the signal from the Ge-C bonds is expected to be much smaller than the Ge-Ge one, the latter signal coming from several Ge layers beyond the surface. The best fit gives an intensity of the Ge-C doublet ~ 10 % with respect to that of the Ge-Ge one. Taking into account the lower atomic surface density of Ge(001) with respect to that of sample A, this value is compatible with about 1 over 5 C atoms of sample A bonded to Ge.

Further information on the nature of the CP phase can be obtained by the comparison between XPS and Raman data. As matter of fact the ratio $R(sp^2)$ between the G peak area of sample A and SLG can be used to estimate the degree of $sp^2$ hybridization in sample A. Indeed, the area of the G peak is proportional to the number of the C-C $sp^2$ bonds [40] and in the SLG all the C bonds are $sp^2$. We found $R(sp^2) = 0.55$, a notably smaller value with respect to the carbon coverage in sample A estimated by XPS suggesting that carbon bonding in the CP phase is not completely hybridized $sp^2$.

*3.2 Scanning tunneling microscopy*

The coexistence of regions exhibiting different carbon bond structures in sample A has been confirmed by STM measurements. The sampling of the surface reveals the presence of two typical morphologies that we identified as characteristic of the CP (Figure 3 (a-e)) and GD (Figure 3 (f-g)) phases. The CP phase morphology is characterized by a flat unfaceted surface (panel (a)) with a rms of about ~0.1 nm and by the presence of round shaped inhomogeneities with typical lateral size of few nanometers (panel (b)). By increasing the magnification (Figure 3(c)) few atom aggregates (C or $CH_x$ species) with different size, shape and local ordering in the same scan area become visible. In panels (*d*) and (*e*) we report magnified areas evidencing the details of these aggregates showing different features. In panel (d) isolated "hexagonal" carbon clusters (lower image) having the same lattice parameter of graphene and linear chains (top image) are shown. Panel (e) reports larger "spot" elements whose internal structure cannot be resolved suggesting a less ordered atomic arrangement. As shown in the profile in panel (e), the typical distance between these "spots" is around 0.3 nm, i.e. basically twice the interatomic distance in graphene.

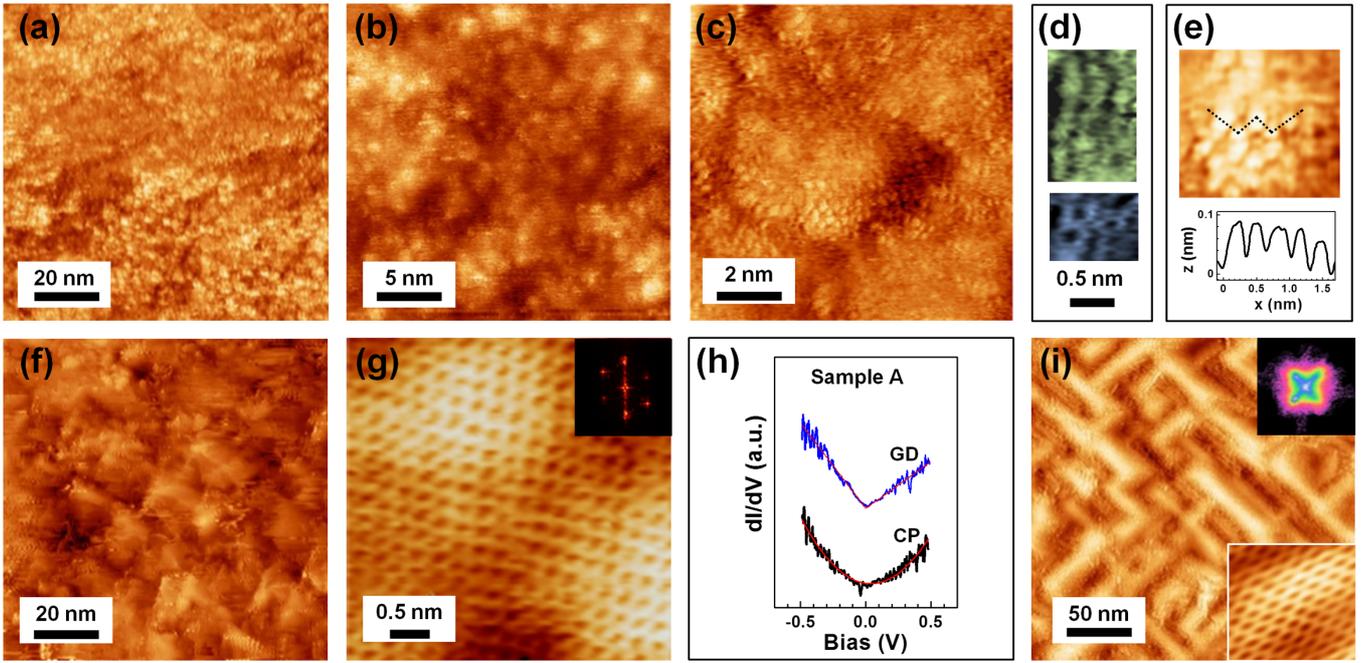

*Figure 3.* STM/STS investigation of sample A and SLG. (a-c): STM images of the CP phase of sample A taken at different magnification. (d-e): details of the C aggregates of CP phase in sample A. In (d) linear C chains (top image) and "hexagonal" carbon clusters (lower image). In (e) disordered "spots" and their profile. (f-g): STM images of the GD phase of sample A, taken at different magnification. In the inset of panel (g) is reported the FFT of the image evidencing the hexagonal symmetry of the lattice. (h) STS measurements: dI/dV curves acquired on CP phase (black curve) and GD phase (blue curve), respectively. In red the quadratic and linear fits of the CP and GD dI/dV curves, respectively. (i) STM image of the SLG sample: the large scale data show the hill and valley nanofaceting. Insets: facet plot (upper) and 2×2 nm$^2$ atomic scale STM image (lower) evidencing the hexagonal structure of graphene. The sides of all the images are along the <110> directions.

As for the GD phase, panel (f) shows a 100 nm×100 nm morphology characterized by the presence of structures whose shape remotely reminds a hut roof. In the following we will refer to the facets of these structures as *proto-facets*. The average height of the *proto-facets* is 0.6 nm while the angles with respect to the surface is ~6-7°. Interestingly their orientation is along the <100> directions which are the same directions of the Ge nanofaceting developing during the SLG growth (see Figure 3(i) where the STM measurements performed on the SLG sample are reported for comparison). Notice that the susceptibility of the Ge(001) surface to form {1,0, L} facets has been observed under a number of conditions [41-44]. The STM image acquired at atomic resolution in the GD region of sample A is reported in Figure 3(g). The typical honeycomb hexagonal lattice of graphene is clearly visible. From local profile and FFT analysis the minimum distance between atoms is evaluated, on average, as 0.14 nm, in excellent agreement with the C-C distance in graphene. All the C structures evidenced by STM cover almost uniformly the surface of sample A with a predominance of CP over GD phase, in agreement with the high value of carbon coverage estimated by XPS and with the relative intensity of their components in the C1s spectrum.

Further analysis on sample A is performed using STM/STS techniques that provide information about the nanoscale electronic properties of the samples [45-46]. STS measurements were carried out at room temperature in a range of 0.5 V around the Fermi level. This spectroscopic characterization has been repeated for both CP and GD phases recognized on the sample A and, for each phase, the *I(V)* curves were averaged over areas of about 5x5 nm$^2$. From the averaged *I(V)* spectra we determined the *dI/dV* curves which are connected to the local density of states. The typical *dI/dV* curves acquired on the two phases (Figure 3(h)) evidence two well distinct profiles. A linear characteristic typical of free standing graphene [47] was obtained on the *proto-faceted* area characterizing the GD phase suggesting a weak interaction between graphene and the Ge substrate [9]. A non linear behavior is found in the CP regions in which graphene is still not well developed. The *dI/dV* curve acquired in these regions is well fitted with a quadratic curve. It is interesting to note that a quadratic like behavior in graphene on Ge system is attributed to the presence of interaction between graphene and Ge substrate [9,12,13].

For the sake of completeness, the STM characterization performed on the sample A was repeated on the SLG. The results are reported in Figure 3 (i). In the bottom inset the atomic resolution image is reported showing a well ordered and regular graphene hexagonal lattice. STS measurements on SLG (not shown) evidence linear *dI/dV* characteristics similar to those found on the GD phase of sample A. On a large length scale the sample morphology is dominated by the Ge nanofaceting, a maze made of faceted structures. The facet plot reported in the inset shows that the orientation of the nanofaceting is along the <001> directions. The intense central spot corresponds to low angle values and takes into account the flat top morphology of faceted structures. The angle that the facets form with the substrate is compatible with {1,0,10}facet orientation in agreement with the previous report based on AFM measurements [10]. Values of the facet angle of nanostructured Ge surfaces between 5° and 8 ° are reported in literature with slight variations of the angles found as a function of the deposition parameters [9, 48].

## 4. Discussion

It was suggested that graphene nucleation occurs when the C adatom concentration on the substrate surface reaches a critical value corresponding to a supersaturation condition [10, 49]. The data acquired on a partial graphene growth performed via CVD (sample A) indicate that this is a relatively slow process in our growth conditions. Indeed, after 30 min of deposition time, graphene domains (GD phase) are present only in very small regions covering about 25% of the surface. Most of the C atoms present on the Ge surface at this stage of the growth are aggregated in the CP phase representing therefore the intermediate path toward graphene nucleation. Precursor structures for the graphene formation including C dimers, rectangle, zigzag and armchair carbon chains were also observed in graphene deposited on Cu(111) [22]. Theoretical calculations have shown that for substrates that are weakly interacting with C atoms, the interaction between two C adatoms is attractive leading to easy ad-dimers formation during graphene growth [15] and that the sp hybridized C-C bond is favored against to $sp^2$ and $sp^3$ C-C bonds [16]. The resulting microscopic path to graphene nucleation on Cu has thus been theoretically identified in linear C chains that transform in Y type and extended polyyne chains with subsequent ring condensation and formation of $sp^2$ hybridized islands [16, 50], species that resemble the C aggregates we identified in our sample A. Our Raman spectroscopy data suggest that in the CP phase carbon bonding is not completely hybridized $sp^2$ and the XPS Ge3d data suggest the presence of Ge-C bonds, instead not present in the SLG. Therefore, the C1s component at 284.1 eV related to the CP phase is likely due to C-C sp bonds in the linear chains, $sp^2$ defected bonds and C atoms bonded to one Ge whose binding energies are all around 284 eV [37,38,51,52], too close each other to be resolved unambiguously. Notice that the presence of C-Ge bonds at the early stage of CVD graphene nucleation on Ge (001) has been predicted theoretically in Ref. [28]. As a matter of fact the DFT calculations of J. Dabrowski at al. suggest that the products of precursor dissociation are strongly bonded to the Ge(001) dimers.

During the graphene nuclei enlargement the GeC bond is substituted by H atoms that saturate the nuclei edge allowing the diffusion of structures made of several carbon rings on the surface during the growth process. The diffusion of carbon clusters on the Ge surface can favor the crystallization process we found as a function of deposition time from sample A to SLG via a Smoluchowski ripening mechanism similar to that observed in the initial stage of graphene growth on metal in Refs. [27,53]. Furthermore during this crystallization process methane can play the dual role of carbon source that maintains the active species concentration above the equilibrium and defect healer as reported in Refs. [22, 50]. Interestingly, in Ref. [50] the authors found that an important role in this defect healing process is played by kinetics effects which are enhanced by the high surface mobility of the catalyst at deposition temperatures close to the substrate melting point, as in our experiment. The mobility and/or evaporation of Ge atoms in our experimental conditions is proved by the evolution of the surface morphology during the graphene growth, from the unstructured flat morphology of CP phase to the proto-faceted one in the GD phase, and finally to the labyrinth faceted structure observed in the SLG. Interestingly, we found that during the growth the nanofaceting develops and expands underneath the graphene film in a compliant manner, with the exposure of the same {1,0,7}-{1,0,10} facets at all the stage of the growth. This finding suggests that although the interaction between graphene and the substrate is weak, it is sufficient to induce and stabilize the faceting in presence of an extremely fluid surface, as already observe in graphene growth on Cu(001) [54].

## 5. Conclusions

In summary, we identified at the early stage of CVD growth of graphene on Ge(001) the carbon precursor phase to graphene nucleation made of C atoms and/or $CH_x$ aggregates partially bonded to the Ge surface. These aggregates do not exhibit a well-established long-range ordering although a local arrangement in linear and hexagonal structures can be recognized in some atomic size regions. The C precursor phase evolves in graphene domains through a crystallization process that in turn results in the formation of a uniform single layer graphene. The nucleation of the small graphene domains is accomplished by the Ge surface *proto-faceting* that evolves in the characteristic Ge nano-faceting of SLG on Ge(001) with the exposure of the same {107}-{1010} facets at all the stage of the crystallization process. This finding suggests that the interaction between graphene and the substrate, although weak, is sufficient to induce and stabilize the faceting in presence of an extremely fluid surface.

**Acknowledgment**

The Lime Laboratory of Roma Tre university is acknowledged for technical support.

Funding from the European Union's Horizon 2020 research and innovation program under grant agreement No. 696656 – GrapheneCore1 is acknowledged.

M.F, L.F., and A.S acknowledge the University of Rome "Tor Vergata" for "Chocolate" project under the grant "Consolidate The Foundations 2015".